\documentstyle[12pt,epsfig]{article} 
\def\bib{\bibitem}
\def\be{\begin{equation}}
\def\ee{\end{equation}}
\def\barr{\begin{array}}
\def\earr{\end{array}}

\def\beq{\begin{eqnarray}}
\def\eeq{\end{eqnarray}}
\def\ra{\rightarrow}
\def\etal{ {\em et al.}}

\def\lsim{\:\raisebox{-0.5ex}{$\stackrel{\textstyle<}{\sim}$}\:}
\def\gsim{\:\raisebox{-0.5ex}{$\stackrel{\textstyle>}{\sim}$}\:}

\def\ra{\rightarrow}

\def\e+e-{$e^+e^-$}
\def\snu{\tilde{\nu}}

\def\ib#1,#2,#3{       {\it ibid.\/ }{\bf #1} (19#2) #3}
\def\ap#1,#2,#3{       {\it Ann.\ Phys.~(NY)\/ }{\bf #1} (19#2) #3}
\def\ijmp#1,#2,#3{     {\it Int.\ J.~Mod.\ Phys.\/ } {\bf A#1} (19#2) #3}
\def\mpl#1,#2,#3 {     {\it Mod.\ Phys.\ Lett.\/ } {\bf A#1} (19#2) #3}
\def\np#1,#2,#3{       {\it Nucl.\ Phys.\/ }{\bf B#1} (19#2) #3}
\def\npps#1,#2,#3{     {\it Nucl.\ Phys.~B (Proc.~Suppl.)\/ }
                          {\bf B#1} (19#2) #3}
\def\plb#1,#2,#3{      {\it Phys.\ Lett.\/ }{\bf B#1} (19#2) #3}
\def\pr#1,#2,#3{       {\it Phys.\ Rev.\/ }{\bf #1} (19#2) #3}
\def\prd#1,#2,#3{      {\it Phys.\ Rev.\/ }{\bf D#1} (19#2) #3}
\def\prep#1,#2,#3{     {\it Phys.\ Rep.\/ }{\bf #1} (19#2) #3}
\def\prl#1,#2,#3{      {\it Phys.\ Rev.\ Lett.\/ }{\bf #1} (19#2) #3}
\def\pro#1,#2,#3{      {\it Prog.\ Theor.\ Phys.\/ }{\bf #1} (19#2) #3}
\def\rmp#1,#2,#3{      {\it Rev.\ Mod.\ Phys.\/ }{\bf #1} (19#2) #3}
\def\sp#1,#2,#3{       {\it Sov.\ Phys.\ Usp.\/ }{\bf #1} (19#2) #3}
\def\zpc#1,#2,#3{      {\it Z. Phys.\/ }{\bf C#1} (19#2) #3}
\def\appb#1,#2,#3{     {\it Acta Phys.\ Polon.\/ }{\bf B#1} (19#2) #3}

\begin{document}

\thispagestyle{empty}
\setcounter{page}{0}
\renewcommand{\thefootnote}{\fnsymbol{footnote}}

% \vspace*{-1in}
\begin{flushright}
DESY 97-050\\[1.7ex]
{\large \tt hep-ph/9703431} \\
\end{flushright}

\vskip 45pt
\begin{center}
{\large \bf Supersymmetric $W$ Boson Decays as a Means to
Search for Charginos and Neutralinos\footnote{Presented at the Cracow 
Epiphany Conference on W Bosons, Cracow, Jan.~1997}
}

\vspace{13mm}
{\large Jan Kalinowski} \\[1.4ex]
{\em Deutsches Elektronen-Sychrotron DESY, D-22607 Hamburg}\\
and\\
{\em Institute of Theoretical Physics, Warsaw University, PL-00681 
Warsaw}\\[2ex]

%\today

\vspace{50pt}
{\bf ABSTRACT}
\end{center}
\begin{quotation}
If the sneutrino mass is below the chargino mass, the dominant decay
mode of the lightest chargino is via a two-body decay channel
$\chi^{\pm}_1 \ra \tilde{\nu} + l^{\pm}$. Sneutrinos are invisible in
$R$-parity conserving supersymmetric models and, if the mass gap
$m(\chi^{\pm}_1)-m(\tilde{\nu})$ is sufficiently small, the soft decay
lepton may escape detection leading to invisible chargino decays.
This ``blind spot'' of the supersymmetry parameter space would
jeopardize the chargino search at LEP2.  We point out that such a
scenario can be tested by searching for single $W$ events in $e^+e^-
\ra W^+W^-$, with one $W$ boson decaying to visible leptons or quark
jets, and the second $W$ boson decaying to invisible charginos and
neutralinos.
\end{quotation}

\def\ra{\rightarrow}
\def\ti{\tilde}

\newpage
\renewcommand{\thefootnote}{\arabic{footnote}}
\section{Introduction}
The $e^+e^-$ collisions at LEP2 energies have greatly improved the
lower mass bounds established at LEP1 on masses of supersymmetric
particles \cite{lep2}, in particular on the lightest chargino mass.
These particles can be produced in pairs in the annihilation process
$e^+e^-\ra \chi^+_1\chi^-_1$ via the $s$-channel $\gamma,Z$ and the
$t$-channel sneutrino $\ti{\nu}_{eL}$ exchanges. The chargino mass
bound depends crucially on the sneutrino mass.  If sneutrinos
$\ti{\nu}_{eL}$ are heavy, the production cross section is large and
the charginos can be probed up to the kinematical limit; only for
small mass gap between chargino and the lightest neutralino, chargino
becomes invisible because the decay fermions (quarks or charged
leptons) in the decay process $\chi^{\pm}\ra \chi^0_1 f\bar{f}'$ are
soft and escape detection.  For $m(\ti{\nu}_{eL})\lsim 200$ GeV, the
destructive interference of $s$- and $t$-channel exchanges reduces the
production cross section, lowering the sensitivity. However, if one of
the sneutrinos is lighter than the chargino by a few GeV, the
sensitivity is lost. In this case, called a ``blind spot'' in
Ref.~\cite{aleph}, the dominant two-body decay mode of the chargino
$\chi^{\pm}_1\ra \ti{\nu}_{lL} l^{\pm}$ is invisible because (a) the
decay lepton $l^{\pm}$ is soft and escapes detection, and (b)
sneutrino is either the lightest supersymmetric particle or it decays
to the lightest neutralino and corresponding neutrino.  Note that the
other two-body decay process, $\chi^{\pm}_1\ra \nu_{lL} \ti{l}^{\pm}$,
due to the SU(2) mass relation\footnote{We consider a low-energy
  supersymmetry with no reference to grand unified scenarios.}
\beq
m^2(\tilde{l}_L)=m^2(\snu_{lL})- m^2_Z\, \cos^2\theta_W \, \cos
2\beta  \label{su2}
\eeq
is closed kinematically because for the sneutrino almost degenerate in
mass with the chargino, $m(\tilde{l}_L)> m(\chi^{\pm}_1)$ for the
preferred values of $\tan\beta>1$ .

The ``blind spot'' is particularly annoying because the charginos could be
as light as 45 GeV, the ultimate limit established at
LEP1~\cite{lep1}. There are several methods to eliminate this
particular region of the parameter space by exploiting: (i)
constraints from future high-precision measurements of $(g-2)_{\mu}$
\cite{carena}; this method works large $\tan\beta\gsim 20$, (ii) the
non-observation of the corresponding left-chiral slepton with the mass
given by Eq.~(\ref{su2}), (iii) single photons in $e^+e^-\ra \gamma
\chi^+_1\chi^-_1$ with charginos undetected; however the production
cross section is small and the background large.

I would like to report on a recent work, done in collaboration with
P.~Zerwas \cite{KZ}, in which we point out that the blind spot can be
explored experimentally by searching for single visible $W$'s in the
$WW$ pair production process $e^+e^-\ra W^+W^-$.  If charginos are as
light as 45 GeV, $W$ bosons can decay invisibly via charginos and
neutralinos, $W^{\pm}\ra \chi^{\pm} \chi^0$. From the measurements of
the total $W$ boson decay width, non-standard $W$ decays are possible
with a branching ratio of $\lsim 7$\% \cite{pdg}. In $WW$ pair
production processes in $e^+e^-$ collisions such invisible
supersymmetric $W$ boson decays in one hemisphere can be tagged by the
observation of the standard decay modes to leptons or quark jets of
the other $W$ boson in the opposite hemisphere\footnote{This is
  similar to the model-independent Higgs boson search in the Bjorken
  process $e^+e^-\ra ZH$ by tagging only $Z$ bosons in the final
  state.}.  We show that for the invisible $W$ decay modes at the
level of a few percent, such processes should be detectable at LEP2
energies.  Their non-observation will allow us to close the region
$m(\chi^{\pm}_1) \gsim m(\tilde{\nu})$ of the parameter space.

\section{Invisible Supersymmetric $W$ Decays}
With LEP1 limits on the supersymmetry parameter space, the decay of
the $W$ bosons to charginos and neutralinos are kinematically open:
\beq 
W^{\pm}\ra \chi^{\pm}_i\chi^0_j \mbox{~~~~~~}[i=1,2;\, j=1,...,4]
       \label{wdec}
\eeq
In practice, it is enough to restrict the analysis to the lightest
chargino in order to allow for maximum phase space. In some areas of
the parameter space the heavier neutralinos $\chi^0_j$ may still be
light enough and their coupling large enough to allow for $W$ decays
into these states too; in the numerical analysis all kinematically
possible decay modes to charginos and neutralinos will be taken into
account.

The supersymmetric $W$ decays have been extensively discussed in the
literature \cite{wdec}. Extending to the case of general mixing in the
chargino and neutralino sectors, the partial widths for the decay
processes (\ref{wdec}) are given by the expression
\beq
&&\Gamma (W^{\pm}\ra \chi^{\pm}_i\chi^0_j)= \frac{G_F m^3_W
  \lambda_{ij}^{1/2}} {6\sqrt{2}\pi}
   \\
  &&\times \left\{ \left[2-\kappa^2_i-\kappa^2_j
    -(\kappa^2_i-\kappa^2_j)^2\right] 
(Q^2_{Lij}+Q^2_{Rij})+
  12\kappa_i \kappa_j\, Q_{Lij}\, Q_{Rij}\right \} \nonumber
\eeq
where $\kappa_i=m_i/m_W$, $\lambda_{ij}=
(1-\kappa^2_i-\kappa^2_j)^2-4\kappa^2_i \kappa^2_j$ is the usual
2-body phase space factor and $m_{i,j}$ are the chargino/neutralino
masses.  The couplings of the $W$ boson to charginos and neutralinos
are written in the usual form as
\beq
Q_{Lij}&=&  Z_{j2} V_{i1} - \frac{1}{\sqrt{2}} Z_{j4} V_{i2}  \\
Q_{Rij}&=&  Z_{j2} U_{i1} + \frac{1}{\sqrt{2}} Z_{j3} U_{i2}
\eeq
where $U$, $V$ are the mixing matrices in the  chargino sector, and 
$Z$ in neutralino sector~\cite{susy}. The mass matrix of charginos
depends on the mixing angle $\beta$ and the wino mass $M_2$;
the neutralino mass depends in addition on the bino mass $M_1$
and the higgsino mass parameter $\mu$. For the sake of
simplicity, in the numerical analysis
below we will adopt the unification  mass relation
$M_1=\frac{5}{3}M_2\tan^2\theta_W$. 

The range of the parameters [$M_2$, $\mu$] for fixed $\tan\beta$ is
restricted by the measurements at LEP1 \cite{lep1}, the non-observation of
of neutrino pair production $\chi^0_1\chi^0_i$ (i=2,3,4) above LEP1
\cite{lep2} and limits on the total $W$ decay width measured at
Tevatron \cite{pdg}. The impact of the AMY limit on $m(\ti{e})\gsim
65$ GeV \cite{amy} is small. The envelope of these constraints, built
up by $m(\chi^+_1)=45$ GeV, $m(\chi^0_1)=12$ GeV and $m(\chi^0_2)=45$
GeV, is shown in Fig.~1 for $\tan\beta=1.5$; the area between and
below the dashed lines is excluded. Note however, that these limits
should only be considered as a guide line because they have not been
derived for the special case $m(\chi^{\pm}_1) \gsim m(\tilde{\nu})$
which is the subject of the present analysis.  For large $\tan\beta$,
the mass limits on charginos and neutralinos forbid on-shell
supersymmetric $W$ decays.

\begin{figure}[htb]
\centerline{\psfig{figure=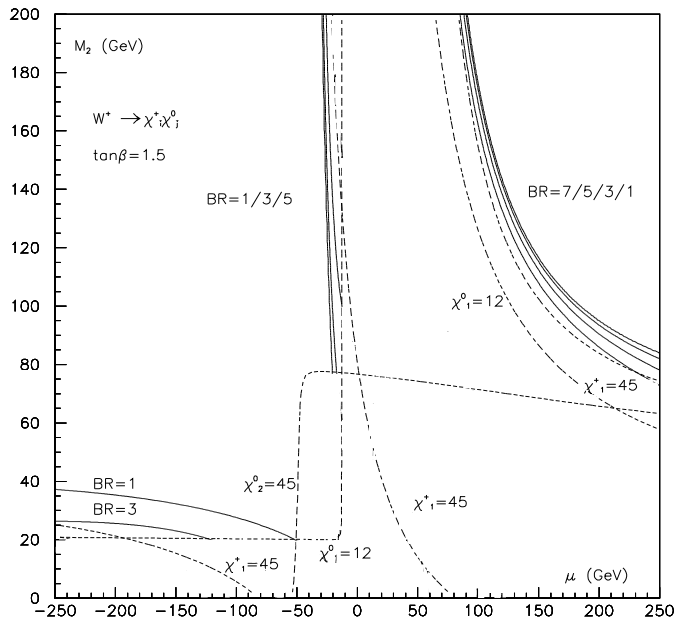,height=10cm,width=10cm}}
\caption{Contour lines for $\tan\beta=1.5$ in the $[\mu,M_2]$ plane 
along which the branching ratios BR$(W\ra \chi\chi)$ of $W$ decays 
to charginos and neutralinos are 7,5,3 and 1\% (full curves). 
Also shown are the contour lines for the mass bounds 
$m(\chi^0_1)=12$ GeV, $m(\chi^0_2)=45$ GeV and $m(\chi^+_1)=45$ GeV 
(dashed curves).}
\end{figure}

In Fig.~1 the solid lines are the contour lines for $W$ decays to
charginos and neutralinos with the branching ratios of 1, 3, 5 and
7\%, with the total decay width given by the standard decay
modes and $\chi^{\pm}\chi^0$ mode. The numbers quoted for the
branching ratios correspond to the partial decay widths of
approximately 20 MeV to 140 MeV. Such decays still can occur in narrow
strips adjacent to LEP1 limits.

The same contour lines for supersymmetric $W$ decays are plotted in
Fig.~2 as a function of lightest chargino and neutralino masses. Only
region $m(\chi^{\pm})>45$ GeV and $m(\chi^0_1)>12$ GeV is shown.  Some
lines terminate in the figure because either $M_2$ or $|\mu|$ is
larger than 400 GeV. In the case of negative $\mu$, the lines
corresponding to 1 and 3\% have two branches, in analogy to Fig.~1.
The cases corresponding to higgsino-like (large $M_2$) and
gaugino-like (large $|\mu|$) light charginos and neutralinos are shown
in the figure. For positive $\mu$ the contour lines extend to
$m(\chi^{\pm}_1)\sim 54$ GeV, for negative $\mu$ up to
$m(\chi^{\pm}_1)\sim 65$ GeV.

\begin{figure}[htb]
\vspace*{-2cm}
\centerline{\hspace{-2cm}\psfig{figure=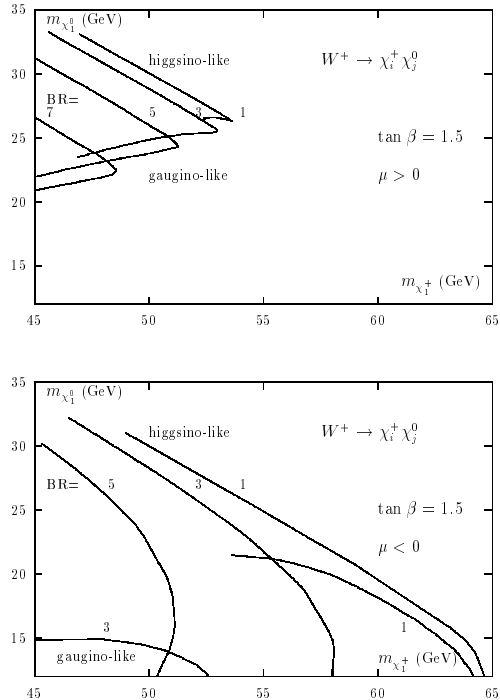,height=17cm,width=12cm}}
\vspace{-4cm}
\caption{Contour lines for $\tan\beta=1.5$, $\mu>0$ [upper plot] 
and $\mu<0$ [lower plot] in the 
$[m_{\chi^+_1},m_{\chi^0_1}]$ plane 
along which the branching ratios BR$(W\ra \chi\chi)$ of $W$ decays 
to charginos and neutralinos are 7,5,3 and 1\%.}
\end{figure}

\section{Tagging Invisible $W$ Decays}
From the Figs.~1 and 2 it is clear that $W\ra\chi\chi$ branching
ratios up to order 7\% are still in the allowed zones of the
$[m(\chi_1^+),m(\chi^0_1)]$ plane.  Assuming a branching ratio of 7\%
for the $W\ra \chi^+\chi^0$ decay modes, one expects the signal
events, defined as one $W$ boson decaying to standard particles and
the other to chargino and neutralino, to occur in 13\% of the cases.
Both $W$ bosons decaying to standard model particles are then expected
in 86.5\% of the cases, and both $W$ bosons decaying to charginos and
neutralinos in 0.5\% of the cases.  Even with limited statistics
collected at the LEP2 measurements so far ($\sim$ 600 $WW$ pairs in
the 4 experiments), we can expect tens of $WW$ signal events events
with mixed standard and supersymmetric $W$ decays. Their observability
depends crucially on the efficiencies for the signal and contamination
from the background processes.

To estimate the feasibility of observing the invisible supersymmetric
$W\ra\chi\chi$ decays in $e^+e^- \ra W^+W^-$ production process, we
consider, as an illustrative example, events collected at the LEP 172
GeV run.  The total $WW$ cross section at this energy is $\sim 13$ pb.
With the combined integrated luminosity ${\cal L} \sim 4\times 11$
pb$^{-1} =44$ pb$^{-1}$ of the four LEP experiments at $\sqrt{s}=172$
GeV, a total of about 570 $WW$ events have been produced, $i.e.$ 1140
$W$ bosons. If $\mbox{BR}(W\ra\chi\chi)=7$\%, the signal cross section
is of the order 1.7 pb, which means that 80 $W$ bosons are potential
candidates for chargino/neutralino decays.  Therefore 74 signal events
with mixed standard and supersymmetric $W$ decays can be expected.

 The signature of these events would be a single $W$ boson, $e^+e^-\ra
W+$ (no other visible particle). They may be tagged in the 2-jet decay
mode or, with reduced branching ratios, in the leptonic $e\nu_e$ and
$\mu\nu_{\mu}$ decay modes. An important feature of these events is
the kinematic constraint that the isolated $W$ bosons carry the beam
energy.  With this kinematical constraint, we expect in the leptonic
tagging mode ($W\ra e\nu /\mu\nu$) an efficiency at least as large as
in the search for acoplanar lepton pairs, $i.e.$ better than 70\%.  In
the 2-jet tagging mode ($W\ra q\bar{q'}$) an efficiency comparable to
that of the search for $WW\ra \tau\nu q\bar{q'}$, $i.e.$ better than
30\% can be achieved. This would give rise to $\sim 10$ signal events
in the leptonic, and $\sim 15$ signal events in the hadronic tagging
mode for the LEP172 run.  If the $\mbox{BR}(W\ra\chi\chi)$ is smaller
than 7\%, the expected number of events is reduced accordingly.

The irreducible background for both the leptonic and 2-jet tagging
modes of the supersymmetric invisible $W$ decays comes from the $WW$
events where one boson decays leptonically with undetected lepton.
Other important background processes include single $W$ final states
$We\nu_e$, and $q\bar{q}\gamma$ events.  In these processes either the
lepton or the photon may escape undetected along the beam pipe giving
rise to a fake ``single $W$'' signal event.  The cross sections for
these background processes have been obtained with the CompHEP program
\cite{boos} without taking into account the hadronization of quarks
and detector effects.  Of course, the hadronization of quark jets and
the smearing due to the experimental resolution must be included when
an experimentally realistic analysis of the signal and background is
performed; however, this is beyond the scope of our analysis.

The background from the $WW$ events is small since only in a small
fraction of the $WW\ra Wl\nu_l$ events the lepton is emitted at a
small angle with the beam pipe. The cross section of 0.03 pb is
expected for events with the lepton in a cone of an half-opening angle
$5^o$ around the beam pipe\footnote{For $WW\ra Wq\bar{q}'$ events due
  to the ``invisible'' SM hadronic decay modes with $q\bar{q}'$
  escaping along the beam pipe, the cross section is of the order 0.02
  pb.}.  The single $W$-boson production is more difficult to
suppress.  An important subprocess in this channel is the
photoproduction process $\gamma e\ra W\nu_e$ with the
Weizs\"acker-Williams photon radiated off the second lepton in the
$e^+e^-$ initial state.  This leads to a background cross section of
0.11 pb and 0.32 pb in the leptonic and 2-jet tagging modes,
respectively. The above cross sections can be further reduced at a
level of 20\% by exploiting the special kinematics of the on-shell
$WW$ signal process, $i.e.$ that the energy $E_i$ of the $W$ decay
products is restricted to the range $ 26\mbox{ GeV} \le E_i \le 62
\mbox{ GeV}$ at $\sqrt{s}=172$ GeV.  The $q\bar{q}\gamma$ final
states, with the photon escaping along the beam pipe, are primarily
induced by the radiative return to the $Z$ with subsequent $q\bar{q}'$
decays, for which a cross section of 120~pb is predicted
\cite{jadach}.  Even though the cross section is large, it can be
suppressed very efficiently by requiring a cut on the invariant mass
of the two jets, 70~GeV~$\le M_{q\bar{q}'}\le$~90~GeV, and the cut on
jet energies, $ 26\mbox{ GeV} \le E_i \le 62 \mbox{ GeV}$, reducing
the value down to 5 pb.  A further cut on the vector sum of the jet
momenta with respect to the beam axis will reduce this background to a
sufficiently low level.

\section{Summary}
If one of the sneutrinos is just below the chargino mass, the
standard experimental search techniques for charginos  in
$e^+e^-\ra \chi^+_1\chi^-_1$  at LEP 2 fail. To probe
this exceptional case we propose to search for ``single $W$''
final states in $WW$ pair production in which one of the $W$
bosons decays invisibly to charginos and neutralinos.  
The special kinematics of on-shell $WW$ production with
2-body $W$ decay, $i.e.$ the invariant mass and the energy constraints,
provide powerful tools to select efficiently
the signal events and to suppress the background processes.
Our estimates of signal and backgrounds show that   
both the leptonic and the 2-jet tagging modes seem to be promising
channels for the search for supersymmetric $W$ boson decays at the
level of a few percent even with the limited statistics
collected so far at LEP 2. With the next run at 184 GeV and larger
luminosities a significant improvement in the sensitivity can be
expected. Therefore the analysis of $W$ production in
$e^+e^-$ collisions can be used to exclude part of the area in the
supersymmetry parameter space in which chargino and sneutrino masses
are nearly degenerate -- or to realize this exceptional case
experimentally. The ``blind spot'' left in the analysis of chargino pair
production in $e^+e^-$ annihilation can thus partly be closed by
exploiting $WW$ production data.

\bigskip
\noindent
{\large \bf Acknowledgments:}

I would like to thank Peter Zerwas for a very enjoyable collaboration
and Dirk Zerwas for numerous discussions on the material presented. 
I am also grateful to  W. de Boer for useful comments. 
This work is partially supported by the KBN 
grant  2~P03B~180~09.

\end{document}